\documentclass[a4paper,11pt]{article}
\usepackage{pos}

\usepackage{subcaption}
\usepackage{physics}
\usepackage{color}
\usepackage{slashed}

\definecolor{bastisnicecolor}{RGB}{248,0,225}

\title{Anomalous transport phenomena on the lattice}

\author[a]{B. B. Brandt}
\author[b]{F. Cuteri}
\author[a]{G. Endr\H{o}di}
\author*[a]{E. Garnacho Velasco}
\author[a]{G. Mark\'{o}}

\affiliation[a]{Universit\"{a}t Bielefeld,\\
Universit\"{a}tsstra{\ss}e 25, 33615 Bielefeld, Germany}

\affiliation[b]{Institute for Theoretical Physics, Goethe University,\\
Max-von-Laue-Stra{\ss}e 1, 60438 Frankfurt, Germany}

\emailAdd{brandt@physik.uni-bielefeld.de}
\emailAdd{endrodi@physik.uni-bielefeld.de}
\emailAdd{cuteri@itp.uni-frankfurt.de}
\emailAdd{egarnacho@physik.uni-bielefeld.de}
\emailAdd{gmarko@physik.uni-bielefeld.de}

\abstract{The interrelation between quantum anomalies and electromagnetic fields leads to a series of non-dissipative transport effects in QCD. In this work we study anomalous transport phenomena with lattice QCD simulations using improved staggered quarks with physical masses in the presence of a background magnetic field. In particular, we focus on the chiral separation effect and calculate the corresponding conductivity both for free quarks and in the interacting case, analyzing the dependence on several parameters, such as the temperature and the quark mass.}

\FullConference{%
  The 39th International Symposium on Lattice Field Theory (Lattice 2022),\\
  8th-13th August 2022,\\
  Bonn, Germany
}

\begin{document}
\maketitle

\section{Introduction} \label{Introduction}

The topological sector of Quantum Chromodynamics (QCD) has been the subject of intense study for more than 40 years, yielding a series of very interesting phenomena associated with it. Some of these effects are called anomalous transport phenomena, as they arise from the interplay between quantum anomalies and electromagnetic fields or vorticities. Topology and quantum anomalies are related via the index theorem, so the manifestation of these non-dissipative transport effects provides an exciting opportunity to probe the non-trivial topological structure of QCD.

One of the most celebrated among these phenomena is the Chiral Magnetic Effect (CME) \cite{Fukushima:2008xe}. The CME has been detected in condensed matter systems \cite{Li:2014bha} and, what is more relevant for our discipline, is actively sought for in heavy ion collision experiments. The most recent result from the STAR collaboration \cite{STAR:2021mii} could not find a signal of CME in a dedicated run in RHIC with isobar collisions. Although the results are still under discussion \cite{Kharzeev:2022hqz}, understanding this apparent suppression of such effects is one of the motivations behind this contribution.  

However, in this work we will focus on another anomalous transport phenomenon, the Chiral Separation Effect (CSE) \cite{Son:2004tq,Metlitski:2005pr}. The objective is to use lattice QCD simulations to determine the CSE conductivity $C_{\text{CSE}}$ and analyze its dependence on relevant parameters, like the temperature $T$ or the mass $m$ of the quarks. This conductivity has been calculated analytically only for free quarks, so the physical setup involving full QCD simulations will certainly help understand how this effect appears in realistic situations. The present effort will also be useful for studying further anomalous transport effects like the CME, for which most lattice simulations so far~\cite{Buividovich:2009wi,Buividovich:2009my,Yamamoto:2011ks,Buividovich:2013hza,Bali:2014vja,Astrakhantsev:2019zkr} were either based on indirect approaches or are yet to be performed in full QCD at the physical point.

This contribution is organized as follows: in section \ref{Anomalous transport phenomena} we review the main features of anomalous transport phenomena. In section \ref{Lattice setup} we present our setup on the lattice and the techniques used to calculate the conductivity. In section \ref{Results for CSE} we present our results for $C_{\text{CSE}}$ both in the free case and in full QCD. Finally, we present our conclusions in section \ref{Summary and Outlook}.

\section{Anomalous transport phenomena} \label{Anomalous transport phenomena}
As it is widely known, QCD is observed to possess CP-symmetry, or equivalently, the $\theta$ parameter is experimentally bound to be practically zero. This fine-tuning issue is known as the strong CP-problem. One of the most interesting characteristics of anomalous transport effects is that they represent a local or event-by-event CP-violation in QCD. 

In this theory, anomalous transport phenomena arise from the $\mathrm{U}_\text{A}(1)$ anomaly. The non-trivial topological solutions of QCD can transfer chirality to quarks through this anomaly, and it is this origin that explains its CP-odd nature. Globally, the expectation value of these anomalous currents has to vanish since the topology has to be trivial in the system with zero topological charge $\qty(\expval{Q_{\text{top}}}=0)$. However, localized regions with non-trivial topology may exist (for example instantons). In these topological environments anomalous phenomena can arise.

\begin{figure}[h]
     \centering
    \includegraphics[width=0.7\textwidth]{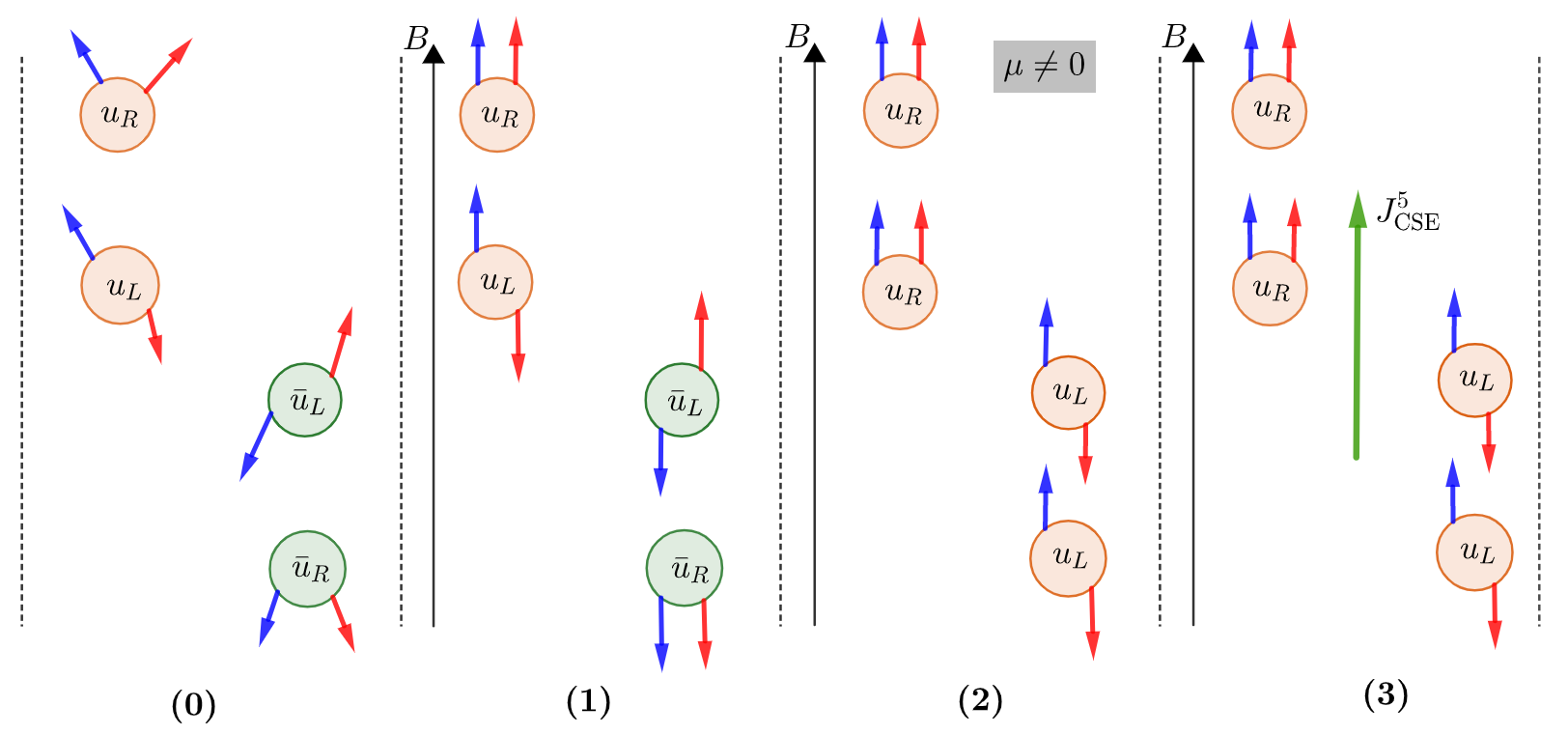}
        \caption{Schematic representation of the CSE in a system with only $u$ and $\Bar{u}$ massless quarks.}
        \label{fig:cse}
\end{figure}

As already mentioned above, the prime example of these effects is the CME: the generation of a vector current in the presence of a chiral imbalance and a magnetic field $B$.
Another example is the CSE, which can be thought of as a ``dual'' of the CME. The CSE is the emergence of an axial current in the presence of finite density and a magnetic field. To first order, the current is linear in the magnetic field (assumed to point in the $z$ direction,) and in the baryon chemical potential $\mu$
\begin{equation}
    \label{eq:cse}
    J_3^5=\sigma_{\text{CSE}}\,  e B=C_{\text{CSE}}\, \mu \, e B + \mathcal{O}(\mu^3)
\end{equation}
with $J_3^5$ the $z$-component of the axial current 
\begin{equation}
    J_3^5=\int \dd^4x \, \Bar{\psi}(x) \gamma_3 \gamma_5 \psi(x).
\end{equation}
This expression holds for a system consisting of a single fermion with charge $e$ and no color. 

Since this is the effect we will be focusing on, let us understand intuitively how this phenomenon works. In Fig.~\ref{fig:cse} we can see a sketch of how the CSE current can appear. Consider a situation with only massless $u,\Bar{u}$ quarks for simplicity (0). When a strong magnetic field is applied (1), all the particles are in the lowest Landau level and their spins align with the direction of $B$ (anti-align if the charge of the particle is negative). At finite density (2), there is a net surplus of particles (supposing $\mu>0$) and right-handed and left-handed particles would flow in different directions, separating particles with different chiralities and thus creating an axial current, which is precisely the CSE current (3).

For non-interacting quarks, an analytical treatment of this problem gives~\cite{Son:2004tq,Metlitski:2005pr}
\begin{equation}
\label{eq:free tot}
     \sigma_{\text{CSE}} =\dfrac{1}{4\pi^2}n_m(T,\mu)
\end{equation}
where
\begin{equation}
    n_m(T,\mu)=\int^\infty_{-\infty} \dd p_3 \qty[n\qty(\sqrt{p_3^2+m^2})+n\qty(-\sqrt{p_3^2+m^2})]
\end{equation}
and $n(E)$ is the usual Fermi-Dirac distribution
\begin{equation}
        n(E)=\dfrac{1}{\exp{(E-\mu)/T}+1}.
\end{equation}
In the case of massless quarks (or, equivalently, asymptotically high temperatures), this expression simplifies to
\begin{equation}
\label{eq:free massless}
     C_{\text{CSE}}=\dfrac{1}{2\pi^2}.
\end{equation}
One of the long-standing questions raised about the CSE (and in general about anomalous transport effects) is how this coefficient is modified by interactions. In the first years after the proposal of anomalous transport phenomena, it was believed that the conductivity did not change in the full theory, as it was topologically protected and hence was not modified by gluonic interactions. However, later works are pointing to corrections of this coefficient in the interacting theory. Since these corrections are non-perturbative, lattice QCD provides an optimal tool to study these effects~\cite{Buividovich:2009wi,Buividovich:2009my,Yamamoto:2011ks,Buividovich:2013hza,Bali:2014vja,Astrakhantsev:2019zkr}. 

\begin{figure}[t]
    \centering
\includegraphics[width=0.5\textwidth]{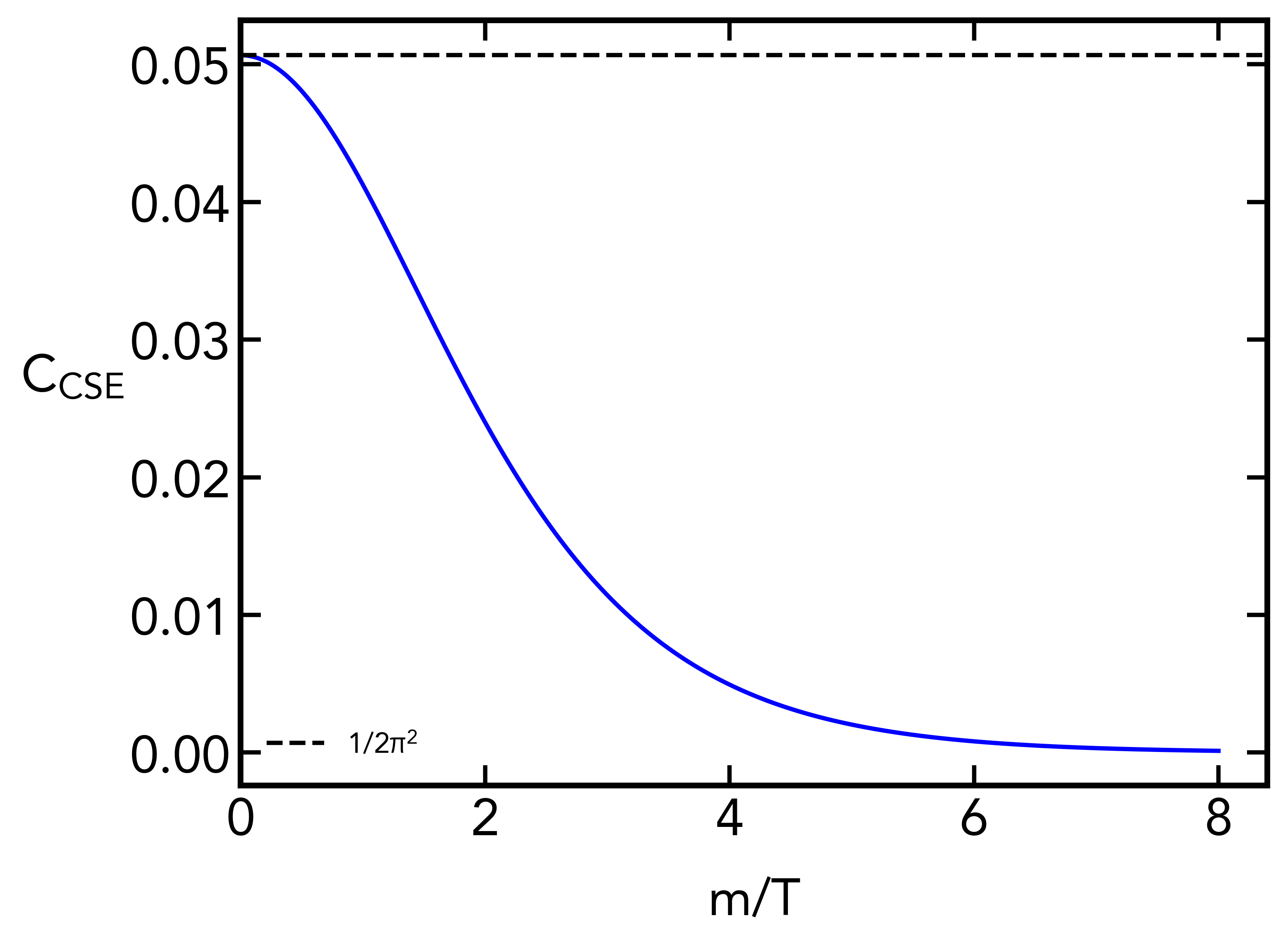}
    \caption{Numerical value of $C_{\text{CSE}}$ as a function of $m/T$ using Eq.~(\ref{eq:free tot}). $C_{\text{CSE}}$ is the numerical coefficient of the first order contribution in $B$ and $\mu$ to the current, so in terms of Eq.~(\ref{eq:free tot}) it can be defined as $\frac{1}{4\pi^2}\eval{\dv{n_m}{\mu}}_{\mu=0}$.} 
        \label{fig:mTint}
\end{figure}

\section{Lattice setup} \label{Lattice setup}
Simulations at finite real $\mu$ suffer from the infamous sign problem. That is why former efforts to calculate $C_{\text{CSE}}$ in the full theory with lattice QCD relied on special setups where this issue does not appear, for example in the quenched theory \cite{Puhr:2016kzp}, where no significant corrections were found, or in two-color QCD \cite{Buividovich:2020gnl}, where at high $T$ the conductivity approached the free case result and for low $T$ CSE was found to be suppressed.

In this work, we take a different approach. Rather than simulating at finite $\mu$ and $B$, we can measure derivatives of the CSE current with respect to the baryon chemical potential. Using Eq.~(\ref{eq:cse}), the derivative yields
\begin{equation}
   \eval{\dv{\expval{J^5_3}}{\mu}}_{\mu=0}= C_{\text{CSE}} \, e B.
\end{equation}
Employing this, leading-order Taylor expansion of the current only requires simulations at $\mu=0$, free of the sign problem.
Then we can take a numerical derivative (linear fit) of the result with respect to $eB$ to obtain $C_{\text{CSE}}$. 

As mentioned above, the last expression holds for a fermion of charge $e$ and no color. For a system with several quark flavors with charge $q_f=\Tilde{q}_f e$ and $N_c$ colors, the formula is modified to
\begin{equation}
   \eval{\dv{\expval{J^5_3}}{\mu}}_{\mu=0}= C_{\text{CSE}} \,N_c\sum_f \Tilde{q}^2_f \, e B \equiv C_{\text{CSE}} C_{\text{dof}} \, e B
\end{equation}

\begin{figure}[h]
    \centering
    \begin{subfigure}{0.4\textwidth}
    \includegraphics[width=\linewidth]{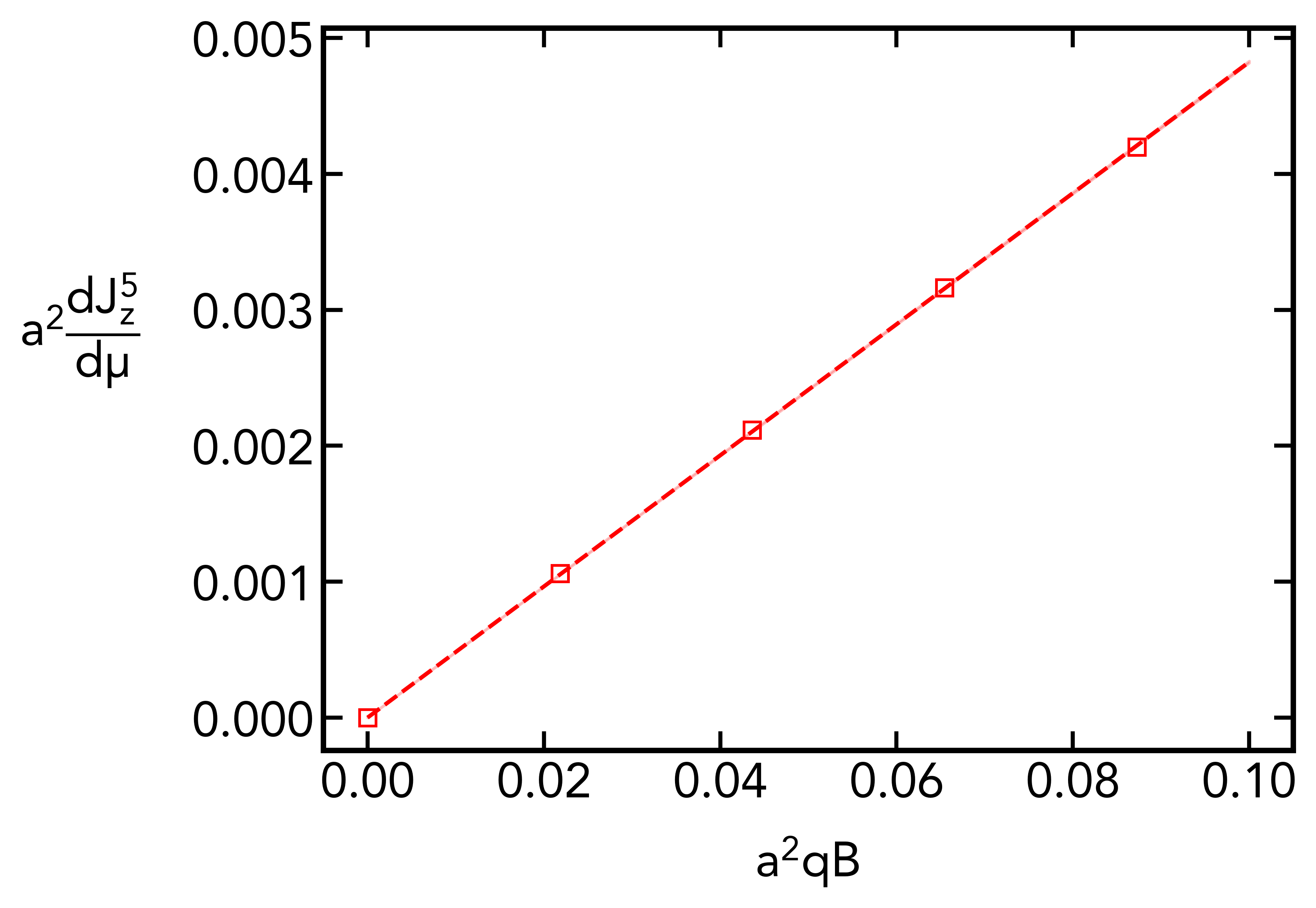}
    \end{subfigure}
    \begin{subfigure}{0.4\textwidth}
   \includegraphics[width=\textwidth]{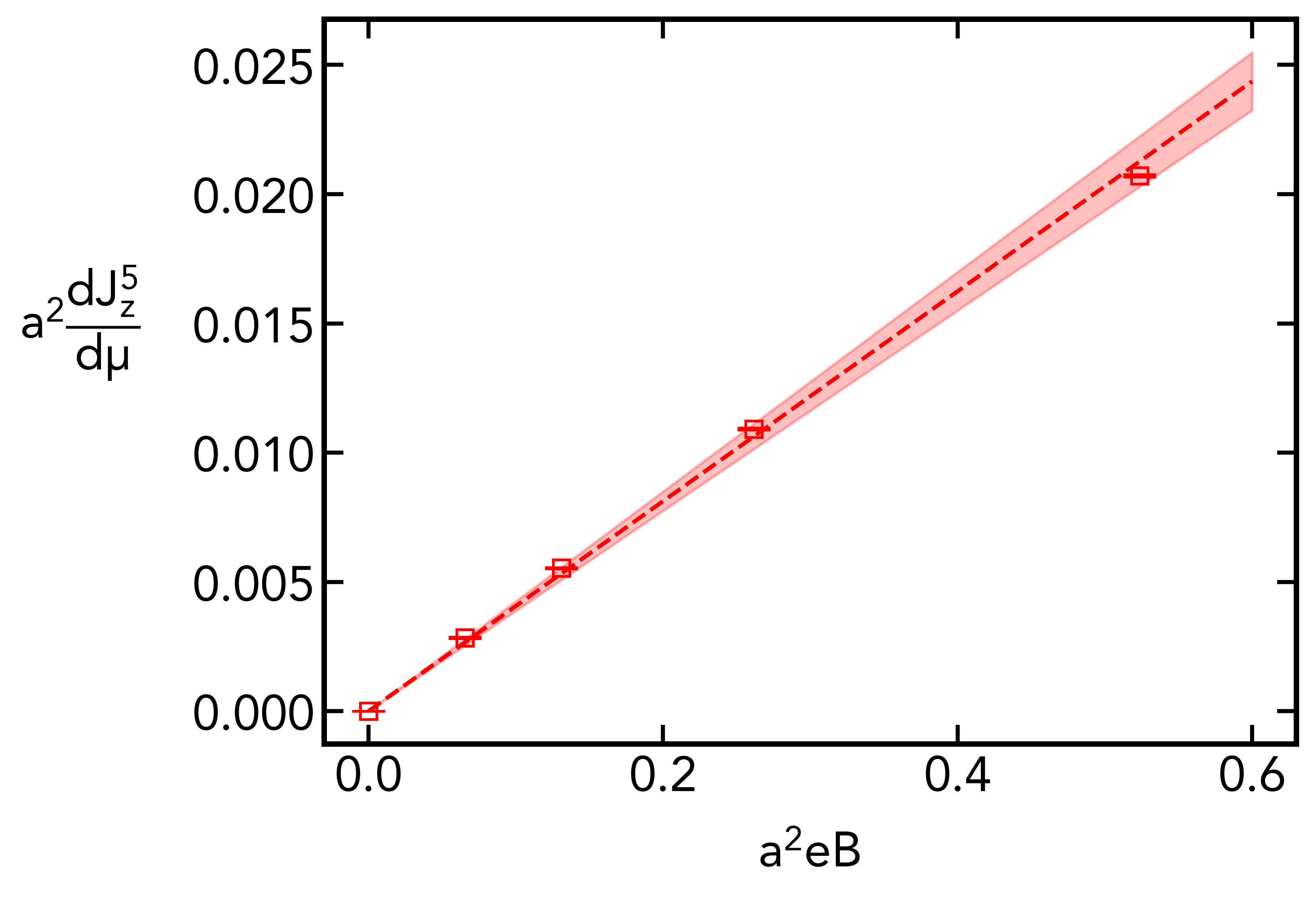}
    \end{subfigure}\\\vspace*{-1mm}
          \caption{Derivative of the CSE (axial) current with respect to $\mu$ as a function of the magnetic field for a $24^3 \times 6$ lattice in the free case (left) and in full QCD with $2+1$ flavors and physical quark masses at $T=305$ MeV (right). As expected, the behavior is linear in both cases and the slope of the fit gives the value of $C_{\text{CSE}}$.}
            \label{fig:fit}
\end{figure}

\noindent with
\begin{equation}
     J_3^5= \sum_f \Tilde{q}_f \int \dd^4x \, \Bar{\psi}_f(x) \gamma_3 \gamma_5 \psi_f(x).
\end{equation}
These are overall factors that can always be restored, so from this point we rescale all our results by $C_{\text{dof}}$.

\section{Results for CSE} \label{Results for CSE}
The measurement of the current derivative in the (rooted) staggered formulation involves a usual connected and a disconnected term, plus an additional term coming from the derivative of the staggered Dirac matrices $\Gamma_\mu$ with respect to $\mu$ 
\begin{equation}
\begin{split}
   \eval{\dv{\expval{ J^5_z}}{\mu}}_{\mu=0}=\dfrac{T}{V}\Bigg[\dfrac{1}{4}&\expval{\text{Tr}\qty(\Gamma_4 M^{-1})\text{Tr}\qty(\Gamma_3\Gamma_5 M^{-1})}_{\mu=0}\\
    -\dfrac{1}{16}&\expval{\text{Tr}\qty(\Gamma_4 M^{-1}\Gamma_3\Gamma_5 M^{-1})}_{\mu=0}\\
    +\dfrac{1}{4}&\expval{\text{Tr}\qty(\dfrac{\partial(\Gamma_3 \Gamma_5)}{\partial \mu}M^{-1})}_{\mu=0}\Bigg]
    \end{split} \label{eq:12}
\end{equation}
where $M=\slashed{D}+m$, $V=(aN_s)^3$ is the spatial volume, $T=(aN_t)^{-1}$ is the temperature, with $N_s$ and $N_t$ the number of spatial and temporal points of the lattice respectively.\footnote{Note that the last term in~\eqref{eq:12} arises because the staggered discretization of the Dirac matrices (see, e.g., Ref.~\cite{Durr:2013gp}) contain links that explicitly depend on the chemical potential.} We have measured the three terms at physical quark masses~\cite{Borsanyi:2010cj} in an already existing ensemble of configurations for different magnetic fields~\cite{Bali:2011qj,Bali:2012zg}. In Fig.~\ref{fig:fit} we show an example of the obtained results. In this plot, we can see that the expected linear behavior is confirmed. However, we are mostly interested in the quantitative result of the slope, so that is what we will be presenting next, both in the free case and in the interacting theory. 

\begin{figure}[h]
    \centering
    \begin{subfigure}{0.4\textwidth}
    \includegraphics[width=\linewidth]{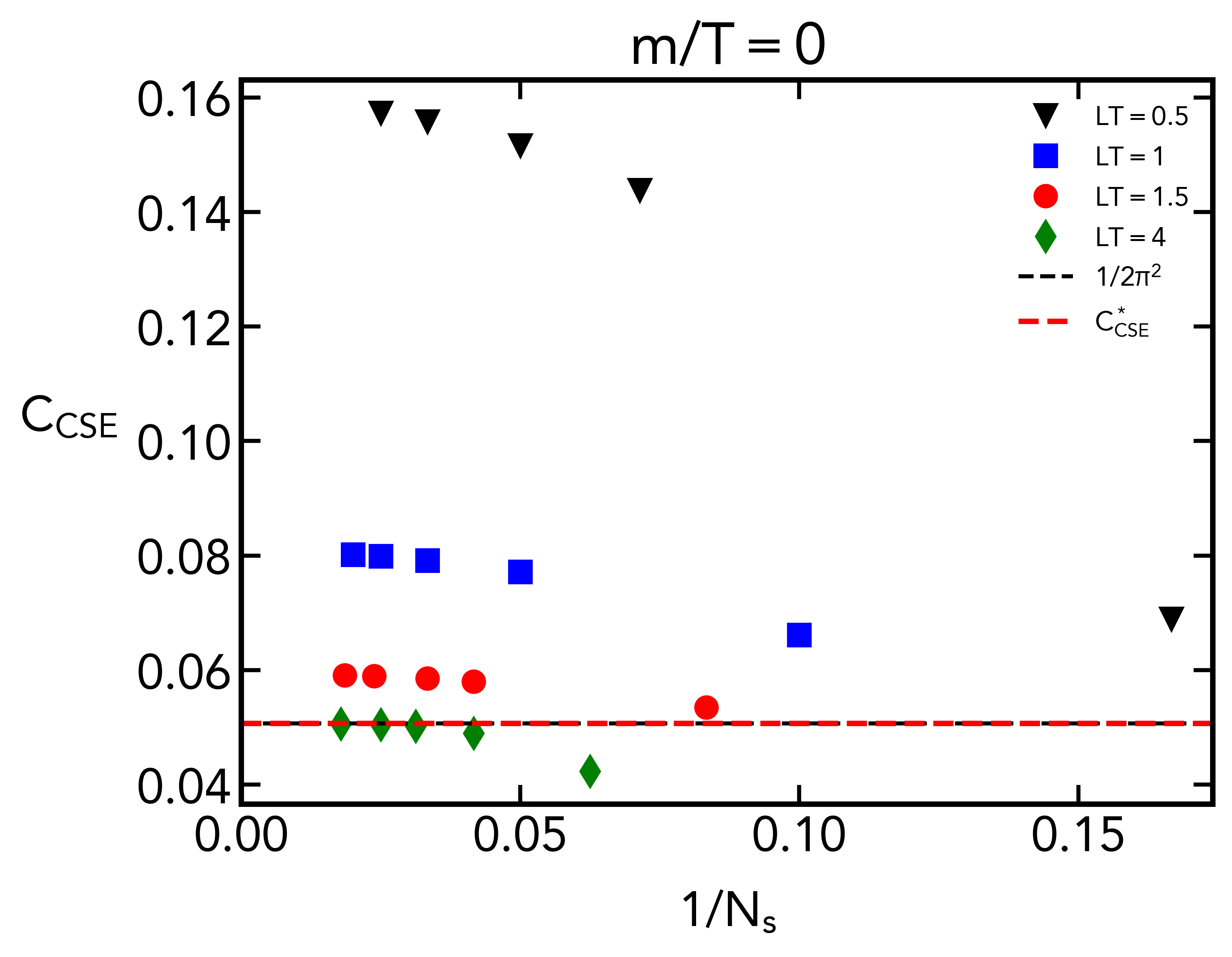}
    \end{subfigure}
    \begin{subfigure}{0.4\textwidth}
    \includegraphics[width=\linewidth]{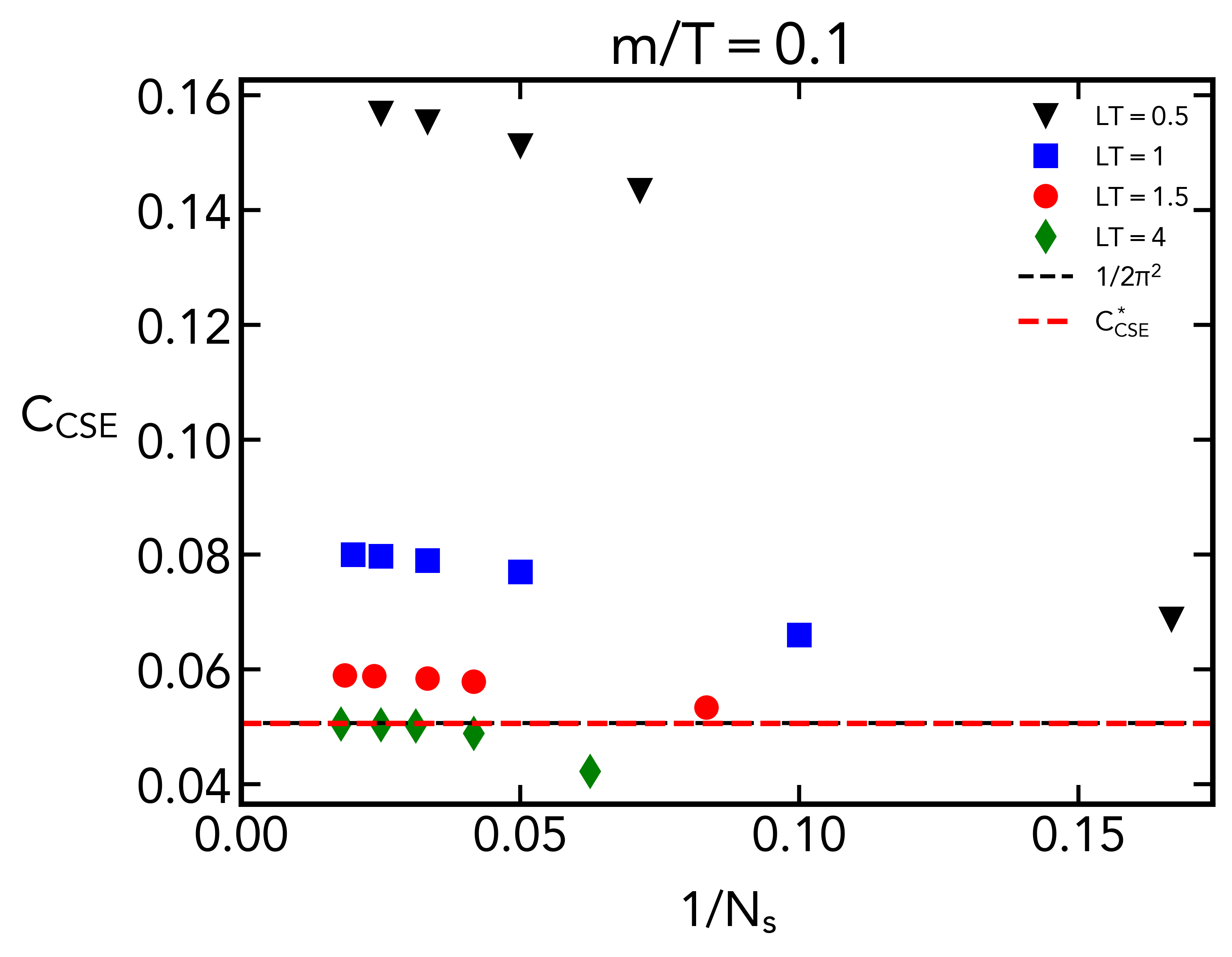}
    \end{subfigure}\\\vspace*{-1mm}
    \begin{subfigure}{0.4\textwidth}
    \includegraphics[width=\linewidth]{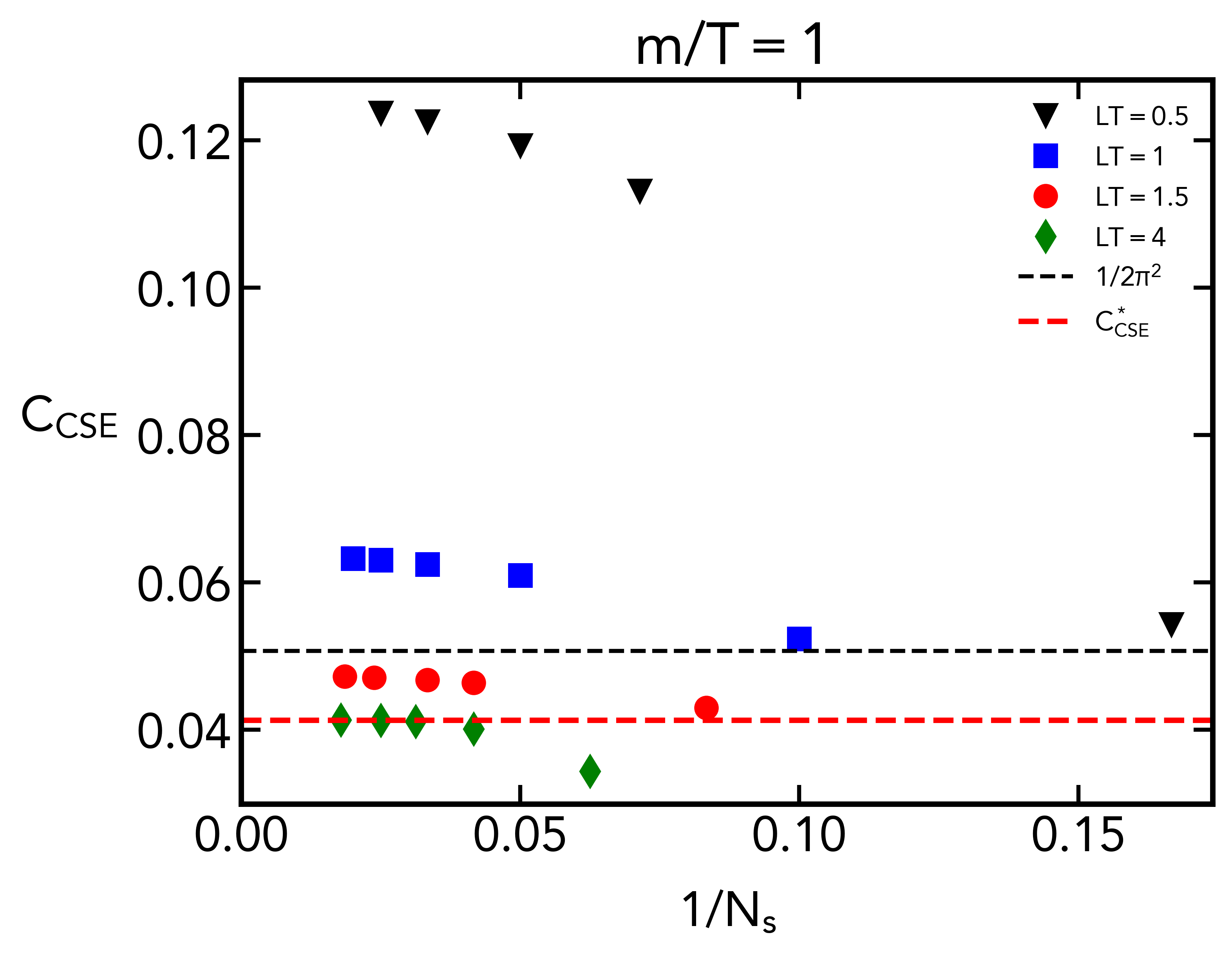}
    \end{subfigure}
    \begin{subfigure}{0.4\textwidth}
    \includegraphics[width=\linewidth]{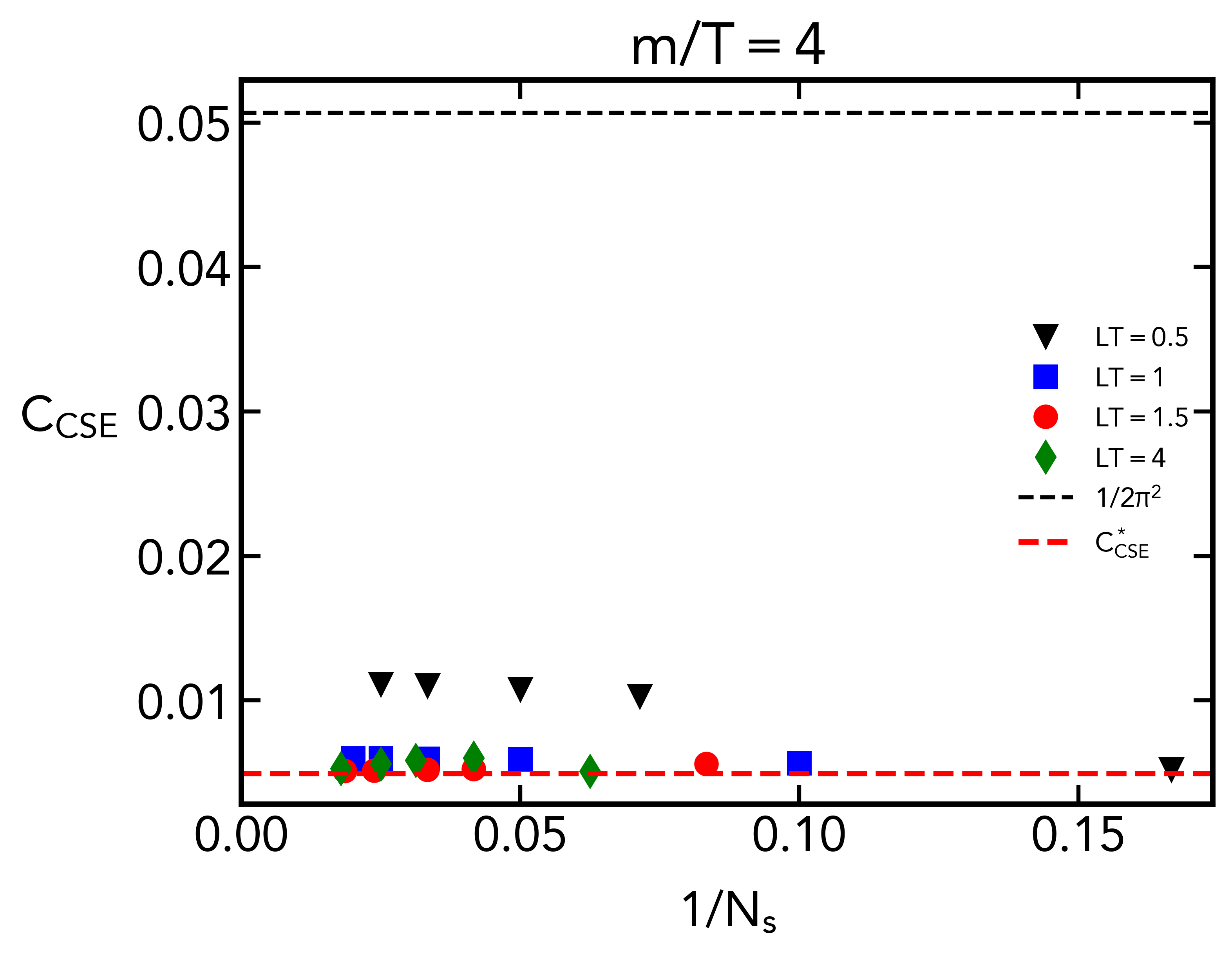}
    \end{subfigure}
    \caption{Results for $C_{\text{CSE}}$ with free staggered quarks at different values of $m/T$ and $LT$. The red dashed line represents the numerical value of $C_{\text{CSE}}$ at the given $m/T$ according to Fig.~\ref{fig:mTint}, while the black dashed line represents $1/2\pi^2$, the value for the analytical calculation of the same coefficient in the massless limit.}
            \label{fig:free}
\end{figure}

Let us start with the free case. In this particular scenario, the sign problem is not present, but we will still use the same approach presented before since we can check the consistency of our setup by comparing to the analytical prediction from Eq.~(\ref{eq:free tot}). In Fig.~\ref{fig:free} we show the free case results for different $m/T$ values, in each plot a continuum limit is taken for different aspect ratios $LT$ by increasing $N_s$ and $N_t$ (with $LT$ and $m/T$ kept constant). For small values of $m/T$, we can see a divergence when $LT$ goes to zero. This behavior disappears when we go towards higher values of $m/T$.  This tells us finite size effects at $LT \rightarrow 0$ are sizeable if $m/T$ is not large enough. It is also worth noting the importance of the continuum limit, since, as can be seen in the plots, non-continuum extrapolated results would underestimate $C_{\text{CSE}}$ if the value of $N_t$ is not large enough. Finally, the main result is that $C_{\text{CSE}}$ approaches the value given by Eq.~(\ref{eq:free tot}) when $LT\rightarrow \infty$ for every value of $m/T$. This serves as a cross-check of our setup since we can reproduce the analytical result for $C_{\text{CSE}}$ with our lattice simulations in the free case.

\begin{figure}[t]
    \centering
\includegraphics[width=0.6\textwidth]{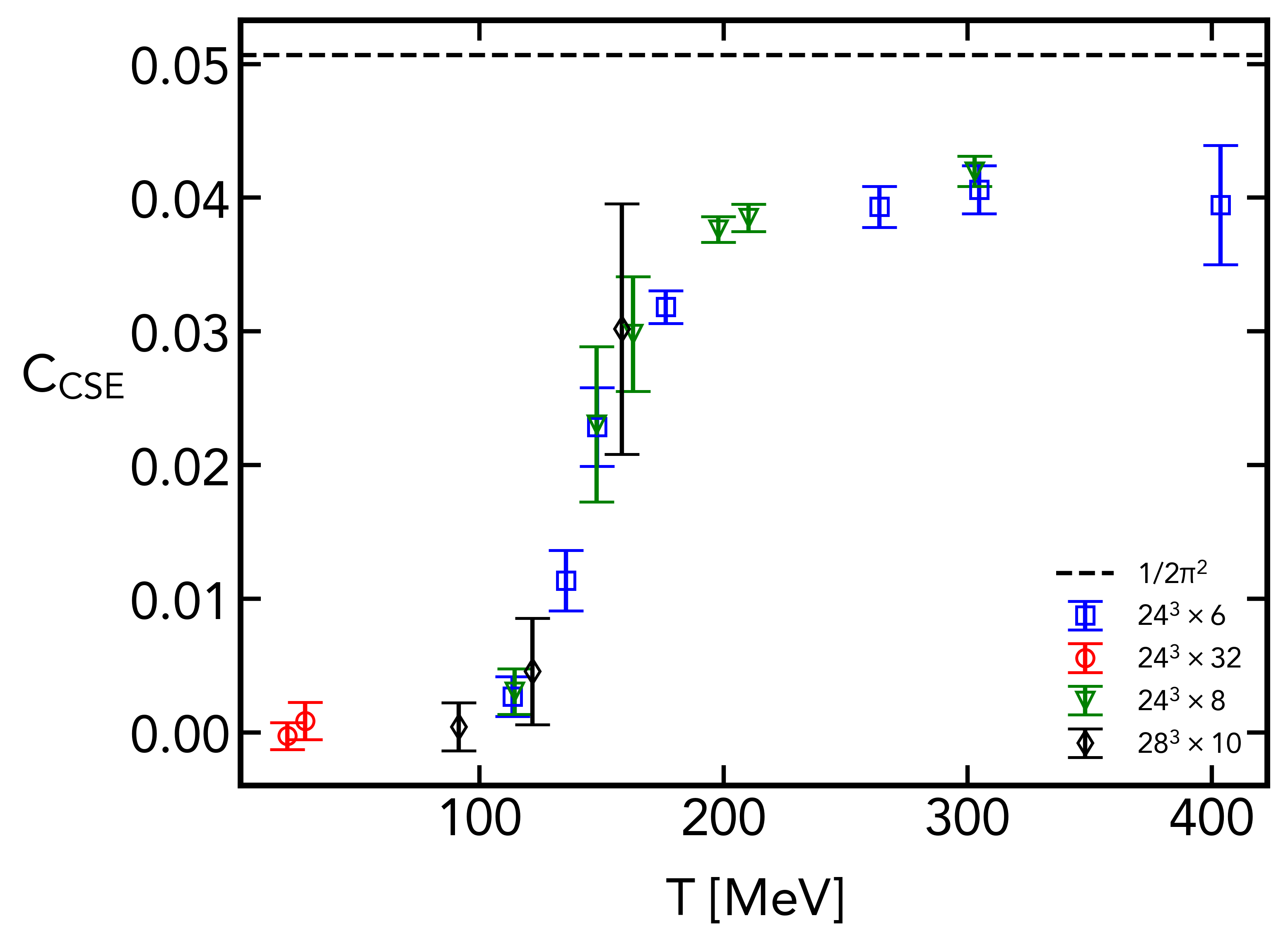}
    \caption{Results for $C_{\text{CSE}}$ with $2+1$ flavors of staggered quarks at physical masses for a wide range of temperatures and four different lattice sizes. The black dashed line represents the analytical prediction for the free case with massless quarks.} 
        \label{fig:full}
\end{figure}

Finally, we present our main result, the conductivity $C_{\text{CSE}}$ in full QCD, in particular for $N_f=2+1$ flavors of staggered fermions at physical quark masses. This is the first fully non-perturbative result for $C_{\text{CSE}}$ at the physical point. Fig.~\ref{fig:full} demonstrates the dependence of the conductivity on the temperature for several finite-temperature lattice ensembles $24^3 \times 6$, $24^3 \times 8$, $28^3 \times 10$ as well as a zero-temperature ensemble $24^3 \times 32$. For temperatures above the QCD transition temperature $T_c$, $C_{\text{CSE}}$ is found to approach the free case prediction. This is in accordance with the expectation based on asymptotic freedom, i.e.\ that at high $T$, QCD approaches a gas of quasi-free quarks and gluons.
At temperatures below $T_c$, the coefficient decreases until reaching zero. This indicates a suppression of the CSE at low temperatures, which is consistent with a previous study in two-color QCD \cite{Buividovich:2020gnl} and can be understood, at a qualitative level, using chiral effective theories \cite{Avdoshkin:2017cqp}. Although a proper continuum and thermodynamic limit is yet to be taken, our results include different lattices spacings and different volumes and the dependence is observed to be minor, so the (qualitative) behavior of $C_{\text{CSE}}$ is not expected to change after taking these limits.

\section{Summary and Outlook} \label{Summary and Outlook}
In this proceedings article, we have presented a study of the CSE using lattice QCD. In particular, we have calculated the conductivity $C_{\text{CSE}}$ using dynamical staggered quarks 
both in the absence of gluonic interactions and in full QCD at the physical point. In the free case, we have recovered the expected value given in Eq.~(\ref{eq:free tot}) once the continuum limit is taken and finite size effects are under control. Having cross-checked our setup, we moved on to full QCD, where we have determined the dependence of $C_{\text{CSE}}$ on the temperature at physical quark masses. At temperatures higher than $T_c$, the coefficient approaches the free case prediction as expected, while at temperatures below $T_c$, the conductivity goes to zero, showing that CSE might be very suppressed at low $T$. This is the first study of this dependence at physical quark masses and it agrees with some of the expectations based on investigations in
two-color QCD. The next steps in this project will be to analyze the mass dependence of $C_{\text{CSE}}$ in the interacting case, as well as to obtain continuum extrapolated results for all the cases. Finally, our technique can be generalized to study the CME, which can contribute to a better theoretical understanding of anomalous transport phenomena, as a counterpart to
the experimental efforts being made to detect this effect.

\acknowledgments
This research was funded by the DFG (Collaborative Research Center CRC-TR 211 ``Strong-interaction matter under
extreme conditions'' - project number 315477589 - TRR 211) and by the Helmholtz Graduate School for Hadron and Ion Research (HGS-HIRe for FAIR). The authors are grateful for inspiring discussions with Pavel Buividovich, Kenji Fukushima, Dirk Rischke, S\"oren Schlichting, Igor Shovkovy and Lorenz von Smekal.  

\bibliographystyle{utphys}
\bibliography{bibliography}

\end{document}